\documentclass[a4paper,12pt]{article}
\usepackage{epsfig}
\usepackage{amssymb}

\newcommand{\be}{\begin{equation}}
\newcommand{\ee}{\end{equation}}
\newcommand{\br}{\begin{eqnarray}}
\newcommand{\er}{\end{eqnarray}}
\newcommand{\ba}{\begin{array}}
\newcommand{\ea}{\end{array}}
\newcommand{\bi}{\begin{itemize}}
\newcommand{\ei}{\end{itemize}}
\newcommand{\bn}{\begin{enumerate}}
\newcommand{\en}{\end{enumerate}}
\newcommand{\bc}{\begin{center}}
\newcommand{\ec}{\end{center}}

\def\OOrd{\buildrel{\scriptscriptstyle >}\over{\scriptscriptstyle\sim}}

\newcommand{\gsim}{\lower.7ex\hbox{$\;\stackrel{\textstyle>}{\sim}\;$}}
\newcommand{\lsim}{\lower.7ex\hbox{$\;\stackrel{\textstyle<}{\sim}\;$}}

\def\sign{\mathrm{sign}}

\def\ar{\to}

\begin{document}
\tolerance=100000
\thispagestyle{empty}
\setcounter{page}{0}

\begin{flushright}
{\rm RAL-TR-1998-081}\\
{\rm  January 1999} \\
\end{flushright}

\vspace*{\fill}

\begin{center}
{\Large \bf Pseudoscalar Higgs production \\[0.25cm]
in association with stop and sbottom pairs \\[0.35cm]
at the LHC in the MSSM}\\[2.cm]

{{\Large A. Dedes} {\large and} {\Large S. Moretti}} \\[3mm]
{\it Rutherford Appleton Laboratory,
Chilton, Didcot, Oxon OX11 0QX, UK} \\[10mm]
\end{center}

\vspace*{\fill}

\begin{abstract}{\small\noindent
We study the processes $gg\ar {\tilde{q}}_1{\tilde{q}}_2^* A$,
with $q=t,b$, within the theoretical framework of the 
Supergravity (SUGRA) inspired Minimal Supersymmetric Standard Model (MSSM)
at the leading order (LO) in perturbative Quantum Chromo-Dynamics (QCD).
Other than constituting a novel production mechanism of the neutral, 
CP-odd Higgs particle, they also allow one to relate the size of
the corresponding production rates to the ratio of the vacuum expectation 
values of the two Higgs fields $\tan\beta$, to 
the trilinear couplings $A_0$ and to the sign of the Higgsino mass
term $\sign(\mu)$. This interplay is made easier by the absence of any 
mixing in the ${\tilde{q}}_1{\tilde{q}}_2 A$ vertices, contrary to
the case of all other Higgs bosons of the theory.
}\end{abstract}

\vspace*{\fill}
\newpage
\setcounter{page}{1}

\section{Introduction and motivation}
\label{sec:intro}

The CP-attribute of the pseudoscalar Higgs boson induces several other
{\sl oddities}
in its behaviours, with respect to all other Higgs scalars of the MSSM, 
that render such a particle a very attractive candidate for phenomenological 
studies. 

For example, in the $q q A$ Feynman rule, where $q$ represents an 
ordinary heavy quark (hereafter, $q=t,b$), there is no dependence 
on the Higgs mixing angle, $\alpha$, 
contrary to the case of the CP-even scalars, $H$ and $h$.
As for the charged Higgs, in the $q\bar q' H^\pm$ vertex, another mixing,
this time in the quark sector, is 
introduced via the Cabibbo-Kobayashi-Maskawa matrix.
A neat consequence of this is the steep rise(fall) of the production cross 
sections of the $A$ boson whenever this is emitted by a heavy down(up)-type
quark, for increasing(decreasing) $\tan\beta(\cot\beta)$ 
\cite{HHG}. For all the other Higgs scalars, such
monotonic behaviour is spoiled 
by the presence of angular terms, typically sines
and cosines of $\alpha$, so that only in extreme regions of the MSSM parameter
space such peculiar dependence on $\beta$ can be recovered. 

If one further considers the Higgs couplings to the scalar partners of
ordinary heavy quarks in Supersymmetric (SUSY) theories, the left- and 
right-handed squarks $\tilde q_{\chi}$, with $\chi=R,L$, then it is easy to 
verify that mixing angles relating their chiral and physical mass eigenvalues
do not  enter the ${\tilde{q}}_1{\tilde{q}}_2 A$ vertex, i.e., the one
involving the observable squarks
(the subscript 1(2) referring to
the lightest(heaviest) of them). Indeed, this is not the case
for the corresponding couplings of  the $H$, $h$ and $H^\pm$ scalars.

One can see this in the context of SUGRA models
\cite{SUGRA}, with the minimal particle content typical of the  MSSM 
(henceforth denoted as M-SUGRA, the environment we choose for our analysis)
\cite{MSSM,MSUGRA}, where the relevant Feynman rules for the 
squark-squark-Higgs vertices can be written in the physical basis 
$\tilde{q}_{1,2}$ as follows \footnote{Note that these Feynman rules 
are valid in a general SUSY model with minimal content independent 
of the origin of the soft breaking terms.}:
%%%%%%%%%%%%%%%%%
\begin{eqnarray}\label{mixing}
\lambda_{\Phi\tilde{q}_1\tilde{q}'_1} &=& c_q c_{q'} \lambda_{\Phi
\tilde{q}_L\tilde{q}'_L} + s_q s_{q'} 
\lambda_{\Phi\tilde{q}_R\tilde{q}'_R}+
c_q s_{q'} \lambda_{\Phi\tilde{q}_L\tilde{q}'_R} + s_q c_{q'} 
\lambda_{\Phi\tilde{q}_R\tilde{q}'_L} ,\nonumber \\[2mm]
%%%%%%%%%%%%%%%%
\lambda_{\Phi\tilde{q}_2\tilde{q}'_2} &=& s_q s_{q'} \lambda_{\Phi
\tilde{q}_L\tilde{q}'_L} + c_q c_{q'} 
\lambda_{\Phi\tilde{q}_R\tilde{q}'_R}
-s_q c_{q'} \lambda_{\Phi\tilde{q}_L\tilde{q}'_R} 
-c_q s_{q'}
\lambda_{\Phi\tilde{q}_R\tilde{q}'_L} ,\nonumber \\[2mm]
%%%%%%%%%%%%%%%%%%
\lambda_{\Phi\tilde{q}_1\tilde{q}'_2} &=& -c_q s_{q'} \lambda_{\Phi
\tilde{q}_L\tilde{q}'_L} + s_q c_{q'} 
\lambda_{\Phi\tilde{q}_R\tilde{q}'_R}
+c_q c_{q'} \lambda_{\Phi\tilde{q}_L\tilde{q}'_R} 
-s_q s_{q'}
\lambda_{\Phi\tilde{q}_R\tilde{q}'_L} ,\nonumber \\[2mm]
%%%%%%%%%%%%%%%%%%
\lambda_{\Phi\tilde{q}_2\tilde{q}'_1} &=& -s_q c_{q'} \lambda_{\Phi
\tilde{q}_L\tilde{q}'_L} + c_q s_{q'} 
\lambda_{\Phi\tilde{q}_R\tilde{q}'_R}
-s_q s_{q'} \lambda_{\Phi\tilde{q}_L\tilde{q}'_R} 
+c_q c_{q'}
\lambda_{\Phi\tilde{q}_R\tilde{q}'_L}. 
%%%%%%%%%%%%%%%%%%
\label{fr}
\end{eqnarray}
%%%%%%%%%%%%%%%
Here, the symbol $\Phi$ denotes cumulatively the five Higgs 
scalars of the MSSM, $\Phi=H,h,A,$ and $H^\pm$.
All the $\lambda_{\Phi\tilde{q}_\chi\tilde{q}'_{\chi'}}$'s appearing in 
eq.~(\ref{mixing}) can be found, e.g., in the Appendix of Ref.~\cite{HHG}.
These are function of the five independent
parameters defining the 
 M-SUGRA model: the universal scalar and gaugino masses $M_0$ and 
$M_{1/2}$, the 
universal trilinear breaking terms $A_0$,  
the ratio of the vacuum expectation values (VEVs) of the two Higgs fields 
$\tan\beta\equiv v/v'$, and the sign of
the Higgsino mass term $\mu$. 

%\footnote{The gaugino mass $M_{1/2}$ does 
%not appear explicitly in eq.~(\ref{mixing}). Further note that 
%the values of the M-SUGRA parameters $M_{0}$, $A_0$, 
%$\tan\beta$ and ${\mbox{sign}}(\mu)$ entering eq.~(\ref{mixing}) 
%are those defined at the 
%electroweak (EW) scale, i.e., after the evolution of the renormalisation 
%group equations (RGEs) starting from the 
%Grand Unification scale $M_{\mathrm{GUT}}$ \cite{graham}.}.

The squarks mixing angles too, i.e., $\theta_q$, with $q=t,b$, 
can  be written in terms of the
above four M-SUGRA parameters, as (here, $s_q\equiv\sin\theta_q$ and
$c_q\equiv\cos\theta_q$)
%%%%%%%%%%%%%%%%%%%%%%
\begin{eqnarray}\label{angle}
\tan(2\theta_b) &=& \frac{2 m_b (A_b + \mu \tan\beta)}{M_{\tilde{Q}_3}^2-
M_{\tilde{D}_3}^2+(-\frac{1}{2}+\frac{2 s_W^2}{3})M_Z^2 \cos2\beta}, \nonumber
\\[2mm]
\tan(2\theta_t) &=& \frac{2 m_t (A_t + \mu \cot\beta)}{M_{\tilde{Q}_3}^2-
M_{\tilde{U}_3}^2+(\frac{1}{2}-\frac{4 s_W^2}{3})M_Z^2 \cos2\beta}, 
\end{eqnarray}
%%%%%%%%%%%%%%%%
with $M_Z$ the $Z$-boson mass and $s^2_W\equiv\sin^2\theta_W$ the (squared)
Weinberg angle, $m_t$ and $m_b$ the top and bottom masses, 
where $A_t$ and $A_b$ are the mentioned (top and bottom)
trilinear couplings at the EW scale, 
while $M_{\tilde{Q}_3}$, $M_{\tilde{U}_3}$ and $M_{\tilde{D}_3}$ are the 
running soft SUSY breaking squark masses of the third generation, as
obtained from starting their evolution at a common 
Grand Unification Theory (GUT) scale set equal to $M_0$. 

Now, it should be noticed  that in the case of the 
CP-odd Higgs boson, i.e., $\Phi=A$,
if one reverts the chirality flow in the vertex $\lambda_{A\tilde{q}_L
\tilde{q}_R}$, the corresponding Feynman rule  changes its sign 
\cite{HHG}\footnote{Indeed, this is the reason why vertices of the
form $A{\tilde q}_{\chi}{\tilde q}_{\chi'}$ with $\chi\equiv\chi'$
are prohibited at tree-level.},
%%%%%%%%%%%%%%%%%%%%
\begin{equation}
\lambda_{A\tilde{q}_L\tilde{q}_R}=-\lambda_{A\tilde{q}_R\tilde{q}_L},
\label{frodd}
\end{equation}
%%%%%%%%%%%%%%%%%%%%%
so that, by making use of eq.~(\ref{fr}) (where 
$\lambda_{A\tilde{q}_1\tilde{q}_1}=\lambda_{A\tilde{q}_2\tilde{q}_2}=0$
\cite{HHG}),
one can conclude that the vertices
$\lambda_{A\tilde{t}_1\tilde{t}_2}$ and 
$\lambda_{A\tilde{b}_1\tilde{b}_2}$ 
are independent of the mixing angles $\theta_t$ and $\theta_b$.
In fact, the Feynman rules for those vertices 
 reduce\footnote{Here, we neglect the CP-violating phases, by assuming
that they have small values, as  preferred 
by the measurements of the Electric Dipole Moments
\cite{nir}. The case in which such phases are sizable 
will be addressed elsewhere \cite{progress}.} 
to (here, $g_W^2=4\pi\alpha_{\mathrm{em}}/s^2_W$ 
and $M_W^2=M_Z^2(1-s^2_W)$)
%%%%%%%%%%%%%%%%%%%%%
\begin{equation}
\lambda_{A\tilde{t}_1\tilde{t}_2} = -\frac{g_W m_t}{2 M_W} \left ( \mu -
A_t \cot\beta \right), \qquad\qquad
\lambda_{A\tilde{b}_1\tilde{b}_2} = -\frac{g_W m_b}{2 M_W} \left ( \mu -
A_b \tan\beta \right ). 
\label{cpoddfr}
\end{equation}
%%%%%%%%%%%%%%%%%%
These are precisely the couplings entering 
the processes that we are going to discuss:
\begin{equation}\label{proc}
gg\ar {\tilde{q}}_1{\tilde{q}}_2^* A,\qquad\qquad
q=t,b.
\end{equation}

It should by now be obvious to the reader our intent in this paper. Namely,
 to study the dependence of the 
production rates of the scattering processes (\ref{proc}) on the 
low-energy SUSY parameters, in order to pin down their actual 
value at the GUT scale, thus constraining the SUSY scenario 
which lies behind the MSSM, through experimental measurements of physical 
observables.

Needless to say, such a task is greatly facilitated in case of pseudoscalar
Higgs production, $\Phi=A$, as we have just shown that the vertex 
expressions for the other cases, when $\Phi=H,h$ and $H^\pm$, are much more 
involved (i.e., they contain as additional free parameters the 
Higgs and squark mixing angles),
so that it is inevitably much more difficult to extract useful 
information from the corresponding production 
rates.
%\footnote{Clearly, other than
%{\sl explicitly} in the Feynman rules, the M-SUGRA parameters also enter 
%{\sl implicitly} in other quantities, such as the scalar masses and
%the running of the strong and EW coupling `constants' (because of the
%structure of the 
%RGEs). Yet,  
%one should expect their main effects to manifest through the 
%triple scalar vertices, as the cross sections typically depend 
%quadratically on them, rather than via suppressed phase space (masses,
%if $\sqrt{\hat{s}}\gg m_{{\tilde q}_\chi} + m_{{\tilde q}_{\chi'}}^* + m_A$) 
%and logarithmic (coupling constants) terms.}.
However, given that the final signatures of all possible 
`gluon gluon $\ar$ squark-squark-Higgs' processes, 
after the decays of the heavy objects,
are at times very similar, one cannot subtract oneself from studying
the whole of such a phenomenology. This is  beyond the scope of this 
short note, though, and we will address the problem 
in a forthcoming publication 
\cite{progress}. Furthermore, in that paper, we will also discuss more closely
another  relevant aspect of Higgs production in 
association with heavy squark pairs, that is, the fact that such processes can
furnish additional production mechanisms of Higgs bosons, to be exploited 
in the quest for such elusive particles, somewhat along the lines of 
Ref.~\cite{abdel}, where the final state with both {\sl light} Higgs, $h$, and 
stop pairs, ${\tilde t}_1{\tilde t}_1^*$, was considered.
Finally, notice that, given the current limits on squark and Higgs
masses \cite{limits,ALEPH}, the only collider environment
able to produce a statistically significative number of events (\ref{proc})
is the LHC ($\sqrt s=14$ TeV), 
to which we confine our analysis. Incidentally, at such a machine, the
contribution to squark-squark-Higgs production via quark-antiquark
annihilations is negligible compared to the gluon-gluon induced rates 
\cite{abdel}, so we will not consider the former here.

The plan of this letter is as follows. The next Section briefly outlines 
how we have performed the calculations. In Sect.~\ref{sec:results} we present
and discuss our findings. The conclusions are in the last Section.

\section{Calculation}
\label{sec:calc}

The techniques adopted to calculate our processes will be described
in detail in Ref.~\cite{progress}, where also the formulae necessary 
for the numerical computation of the Feynman amplitudes will be given. Here, 
we only  sketch the procedure, for completeness.

There are 10 LO Feynman diagrams for each
of  the two processes (\ref{proc}): see Fig.~1 of \cite{abdel} for the
relevant topologies. These have been calculated analytically 
 and integrated numerically over a three-body phase space.
While doing so, they have been convoluted with
gluon Parton Distribution Functions (PDFs), as provided by the LO set 
CTEQ(4L) \cite{CTEQ4}\footnote{The systematic error due the gluon behaviour
inside the proton has been investigated by comparing the CTEQ rates
with those obtained by using other LO fits, such as 
MRS-LO(09A,10A,01A,07A) \cite{MRS98LO}. Typical differences
were found to be less than 15-20\% \cite{progress}.}.

The centre-of-mass (CM) energy at 
the partonic level was the scale used to
evaluate both the PDFs and the strong coupling constant, 
$\alpha_{\mathrm{s}}$. We have used the two-loop scaling for the latter,
with all relevant thresholds \cite{sakis} onset within the MSSM
(as these are spanned through the $Q^2$ evolution of the structure
functions), in order to match the procedure we have adopted in generating the  
other couplings, these also produced via the two-loop RGEs \cite{isajet}.

Depending on the relative value of the final state masses in (\ref{proc}), 
whether $m_{{\tilde{q}}_{2}}$ is larger or smaller than 
$(m_{{\tilde{q}}_{1}} + m_{A})$,
the production of the pseudoscalar Higgs boson can be regarded
as taking place either via a decay or a bremsstrahlung channel. 
We have treated the two processes on the same footing, without
making any attempt to separate them, as
for the time being we are only interested in the total production rates
of the 2 $\ar$ 3 processes (\ref{proc}). 
In this respect, it should be mentioned that the 
partial widths entering in our MEs are
significantly smaller than the total decay widths, so that processes
(\ref{proc}) do retain the
dynamics of the squark-squark-Higgs production vertices also at decay 
level\footnote{Generally, the dominant decay channels of squarks are
to the lightest neutralino/chargino and the `parent' quarks
(see Tab.~V later on).}.

Regarding  the 
numerical values of the M-SUGRA parameters adopted in this paper,
we have proceeded as follows. For a start, we have set 
$M_0=200$ GeV and $M_{1/2}=100$ GeV. For such a  choice, 
the M-SUGRA model predicts squark and 
Higgs masses in the region of 100--400 GeV, so that the latter can
in principle materialise at LHC energies. 
Then, we have varied the trilinear soft SUSY
breaking parameter $A_0$ in a large region, $(-500,500)$ GeV, while we 
have spanned the $\tan\beta$ value between 2 and 45.  
As for $\mu$, whereas in our model its magnitude is constrained, its
sign is not. Thus, in all generality, we have explored both the possibilities
$\sign(\mu)=\pm$. Finally, we have gone back to consider
$M_0$ and $M_{1/2}$ in other mass regions.

{Starting from the five M-SUGRA parameters  
$M_0$, $M_{1/2}$,  $A_0$,  $\tan\beta$ and
${\mbox{sign}}(\mu)$, we have 
generated the spectrum of masses, widths, couplings
and  
mixings relative to squarks and Higgs particles entering reactions (\ref{proc})
by running the {\tt ISASUGRA/ISASUSY} programs for M-SUGRA contained in the 
latest release of the package {\tt ISAJET} \cite{isajet}. 
The default value of the top mass we have used was 175 GeV. Finally,
note that also typical EW parameters, such as $\alpha_{\mathrm{em}}$ and 
$\sin^2\theta_W$, are taken from this program. 

\section{Results}
\label{sec:results}

Although we will in this letter mainly concentrate on the case 
of CP-odd Higgs production, we nonetheless ought to display
some typical cross sections  for all 
processes of the form \cite{progress}
%%%%%%%%%%%%%%%%%
\begin{equation}\label{allproc}
g~g~\rightarrow~
{\tilde{q}}_{\chi}~ {\tilde{q}}^{'*}_{\chi'}~ \Phi,
\end{equation}
where $q^{(')}=t,b$, $\chi^{(')}=1,2$ and $\Phi=H,h,A,H^\pm$. 
This is done in Tab.~I, 
where, for reference, the trilinear coupling $A_0$ has been
set to zero and two extremes values of $\tan\beta$, i.e., 
$2$ and $40$, have been
selected. The corresponding mass spectrum for the particles in the final state
of processes (\ref{allproc}) is given in Tab.~II. There,
it is well worth noticing that modifying the value of $\tan\beta$
corresponds to induce quite different mass values
for both Higgs bosons and squarks. 

From Tab.~I\footnote{Note that, here and in the following, the production
rates do {\sl not} include charge conjugation.}, 
one can notice that our two processes could well yield detectable rates
in the large $\tan\beta$ region. For a LHC running at high luminosity, some
seven thousand such events can be produced per year. 
For large $\tan\beta$
values, alongside pseudoscalar
Higgs boson production, there are at least three other mechanisms
 (\ref{allproc}) with observable rates, as one finds that,
typically:
\begin{equation}\label{large}
\sigma ( gg\rightarrow \tilde{t}_1 \tilde{t}_2^* H;~ 
         gg\rightarrow \tilde{t}_1 \tilde{t}_1^* h;~
         gg\rightarrow \tilde{t}_1 \tilde{t}_2^* h;~ 
         gg\rightarrow \tilde{t}_1 \tilde{t}_2^* A;~ 
         gg\rightarrow \tilde{b}_1 \tilde{b}_2^* A) \gsim 10^{-2} \; 
{\rm pb}.
\label{allcross}
\end{equation}
Among these, it is  $gg\rightarrow \tilde{t}_1 \tilde{t}_2^* H $
that shows a strong dependence on $\sign(\mu)$, whereas all other reactions 
in (\ref{large}) are rather stable against variations of the latter.  
\begin{center}
%%%%%%%%%%%%%%%%%%%%%%%%%%%%%
\begin{eqnarray}
\begin{array}{|c|c|c|c|c|}\hline
\sigma \; {\mathrm{(pb)}}
 & \tan\beta=2,\mu>0 & \tan\beta=2,\mu<0 & \tan\beta=40,\mu>0 & \tan\beta=40,
\mu<0 \\[2mm] \hline
\sigma (gg\rightarrow \tilde{t}_1  \tilde{t}_1^* H ) & 3.3 \times 10^{-4}&
3.6\times 10^{-5} & 4.7\times 10^{-5} & 5.1\times 10^{-5} \\[2mm] \hline
\sigma (gg\rightarrow \tilde{t}_2  \tilde{t}_2^*  H ) & 2.1\times 10^{-5}&
1.9\times 10^{-5}&5.2\times 10^{-6}&6.2\times 10^{-6} \\[2mm] \hline
\sigma (gg\rightarrow \tilde{t}_1  \tilde{t}_2^*  H ) & 4.6\times 10^{-4}&
 2.6\times 10^{-5}& 1.2\times 10^{-2}& 4.6 \times 10^{-3} \\[2mm] \hline
\sigma (gg\rightarrow \tilde{b}_1  \tilde{b}_1^*  H ) &3.0\times 10^{-6}
& 2.8\times 10^{-6} & 2.0\times 10^{-6} & 1.7 \times 10^{-6} \\[2mm] \hline
\sigma (gg\rightarrow \tilde{b}_2  \tilde{b}_2^*  H ) & 5.1\times 10^{-8}&
4.9\times 10^{-8}& 1.0\times 10^{-6} & 1.1\times 10^{-6} \\[2mm]\hline
\sigma (gg\rightarrow \tilde{b}_1  \tilde{b}_2^*  H ) & 1.3\times 10^{-6}&
1.9\times 10^{-7}& 4.1\times 10^{-3}&3.1\times 10^{-3} \\[2mm] \hline
\sigma (gg\rightarrow \tilde{t}_1  \tilde{t}_1^*  h ) & 2.2\times 10^{-1} &
 1.8\times 10^{-2} &
1.4\times 10^{-2} & 1.4\times 10^{-2} \\[2mm] \hline
\sigma (gg\rightarrow \tilde{t}_2  \tilde{t}_2^*  h ) & 3.6\times 10^{-3} &
5.5\times 10^{-3}&
1.4\times 10^{-3}& 1.5\times 10^{-3} \\[2mm] \hline
\sigma (gg\rightarrow \tilde{t}_1  \tilde{t}_2^*  h ) & 1.7\times 10^{-2} &
2.8\times 10^{-2} &7.4\times 10^{-2} & 9.3\times 10^{-2}\\[2mm] \hline
\sigma (gg\rightarrow \tilde{b}_1  \tilde{b}_1^*  h ) & 1.2\times 10^{-4} &
1.5\times 10^{-4} & 2.0\times 10^{-4}& 1.8\times 10^{-4} \\[2mm] \hline
\sigma (gg\rightarrow \tilde{b}_2  \tilde{b}_2^*  h ) & 2.0\times 10^{-6} &
 2.6\times 10^{-6} &
1.4\times 10^{-6} & 1.5\times 10^{-6} \\[2mm] \hline
\sigma (gg\rightarrow \tilde{b}_1  \tilde{b}_2^*  h ) & 3.4\times 10^{-6} &
1.2\times 10^{-4} & 2.4\times 10^{-3} & 3.3\times 10^{-3} \\[2mm] \hline
\sigma (gg\rightarrow \tilde{t}_1  \tilde{t}_2^*  A ) & 5.7\times 10^{-4} &
7.7\times 10^{-5} & 2.9\times 10^{-2} & 1.7\times 10^{-2} \\[2mm] \hline
\sigma (gg\rightarrow \tilde{b}_1  \tilde{b}_2^*  A ) & 1.3 \times 10^{-6} &
2.4\times 10^{-7} & 1.3\times 10^{-2} & 1.2\times 10^{-2}\\[2mm] \hline
\sigma (gg\rightarrow \tilde{t}_1  \tilde{b}_1^*  H^- ) & 4.3\times 10^{-5} &
1.4\times 10^{-5} & 5.0\times 10^{-6} & 4.8\times 10^{-6} \\[2mm] \hline
\sigma (gg\rightarrow \tilde{t}_2  \tilde{b}_2^*  H^- ) & 2.7\times 10^{-7} &
 2.5\times 10^{-7}
& 5.0\times 10^{-4} & 5.4\times 10^{-4} \\[2mm] \hline
\sigma (gg\rightarrow \tilde{t}_1  \tilde{b}_2^*  H^- ) & 5.6\times 10^{-6} &
 3.2\times 10^{-7}
& 4.5\times 10^{-3} & 4.2\times 10^{-3} \\[2mm] \hline
\sigma (gg\rightarrow \tilde{t}_2  \tilde{b}_1^*  H^- ) & 3.7\times 10^{-4} &
 7.9\times 10^{-5} & 2.1\times 10^{-3} & 1.7\times 10^{-3} \\[2mm]
\hline
\end{array}
\nonumber
\end{eqnarray} 
%%%%%%%%%%%%%%%%
{Table I: Total cross sections for processes
of the type 
$g~g~\rightarrow~
{\tilde{q}}_{\chi}~ {\tilde{q}}^{'*}_{\chi'}~ \Phi$,
where $q^{(')}=b,t$, $\chi^{(')}=1,2$ and $\Phi=H,h,A,H^\pm$, in  the
MSSM, at the leading order in perturbative QCD,
for selected values of $\tan\beta$ and $\sign(\mu)$. 
The other three independent
parameters of the model have been set as: $M_0=200$ GeV, $M_{1/2}=100$
GeV and $A_0=0$.}
\end{center}

At low $\tan\beta$'s, the 
cross sections for pseudoscalar Higgs boson production are presumably 
too poor to be of great 
experimental help, in both cases of sbottom and stop squark production.
Even assuming 100 fb$^{-1}$ of 
accumulated luminosity per year at the LHC, only a handful of events of the
form (\ref{proc}) can be produced, if $A_0\approx0$, mainly through the 
$gg\rightarrow \tilde{t}_1 \tilde{t}_2^* A$ channel. (Prospects are somewhat
more optimistic if the common
trilinear coupling is much smaller than zero, 
but in the case of stop squarks only: we will 
come back to this point later on.) 
Anyhow, for small $\tan\beta$ values, the three channels 
$gg\rightarrow \tilde{t}_\chi \tilde{t}_{\chi'}^* h$ 
 can boast very large production rates. Moreover, the dominant
one, when $\chi=\chi'=1$, exhibits
a strong sensitivity on the sign of $\mu$, as the two cross sections
obtained for $\sign(\mu)=\pm$ differ by
about an order of magnitude. 

Therefore, both in the low and high $\tan\beta$ regime, pseudoscalar
Higgs boson production is in general less effective than other channels in 
constraining the sign of the Higgsino mass term. This aspect will 
however not be investigated any 
further here, as it will be addressed in detail 
in forthcoming Ref.~\cite{progress}. In fact, here we are more concerned with
the fact that reactions
(\ref{proc}) are very sensitive to $\tan\beta$,  much more
than any other squark-squark-Higgs production channel: compare the
first two columns in Tab.~I with the last two, particularly for 
$gg\ar \tilde{b}_1 \tilde{b}_2^* A$ (differences are 
over four/five orders of magnitude !).
Other competitive mechanisms in this respect are (in the observable region,
say, with a cross section above $10^{-3}$ pb): 
$gg\ar \tilde{t}_1 \tilde{t}_2^* H$, 
$gg\ar \tilde{b}_1 \tilde{b}_2^* H$  and
$gg\ar \tilde{t}_1 \tilde{b}_2^* H^-$. They are however suppressed, in 
general,  as compared to the two CP-odd 
production channels,
so for the time being we leave them aside for future studies \cite{progress}
and concentrate exclusively on processes (\ref{proc}).

\begin{center}
%%%%%%%%%%%%%%%%%%%%%%%%%%%%%
\begin{eqnarray}
\begin{array}{|c|c|c|c|c|}\hline
{\rm Masses } \; ({\mathrm{GeV}})
 & \tan\beta=2,\mu>0 & \tan\beta=2,\mu<0 & \tan\beta=40,\mu>0 & \tan\beta=40,
\mu<0 \\[2mm] \hline
m_{\tilde{t}_1} & 171   & 258   & 221   & 224 \\[2mm] \hline
m_{\tilde{t}_2} & 374   & 320   & 349   & 348 \\[2mm] \hline
m_{\tilde{b}_1} & 275   & 275   & 222   & 226 \\[2mm] \hline
m_{\tilde{b}_2} & 314   & 314   & 306   & 303 \\[2mm] \hline
m_H             & 382   & 385   & 128   & 126 \\[2mm] \hline
m_h             & 81    & 69^*  & 104   & 103 \\[2mm] \hline
m_A             & 375   & 377   & 128   & 126 \\[2mm] \hline
m_{H^\pm}       & 383   & 385   & 155   & 153 \\[2mm] \hline
\end{array}
\nonumber
\end{eqnarray}
%%%%%%%%%%%%%%%%%%%%%%%
{Table II: Masses of stop and sbottom squarks and   
Higgs bosons
in the MSSM in the small and large $\tan\beta$ region, for both positive
and negative values of $\mu$  and 
universal boundary conditions: $M_0=200$ GeV, $M_{1/2}$=100
GeV and $A_0=0$.}
\vskip0.25in\noindent
$^*$ \footnotesize{This value is actually excluded from the latest LEP
bounds on the light and CP-odd Higgs boson masses \cite{ALEPH}:
$m_h, m_A \gsim 80$ GeV. Nonetheless we keep it here for illustrative
purposes, as representative of the condition $m_A\gg m_h$, typical
of the small $\tan\beta$ regime.}
\end{center}

Figs.~1 and 2 further enlighten the $\tan\beta$ dependence of pseudoscalar
Higgs boson production, in association with stop and sbottom squarks,
respectively, as we have now treated $A_0$ as a variable parameter.
Indeed, the typical behaviour seen in Tab.~I for $A_0=0$ in
$gg\ar \tilde{t}_1 \tilde{t}_2^* A$ and
$gg\ar \tilde{b}_1 \tilde{b}_2^* A$ persists for all other values of $A_0$  
considered. 

The variation with $\tan\beta$, and particularly
the steep rise at high values of the latter, can  be 
understood in the following terms. For large 
$\tan\beta$,  the squark-squark Higgs boson
couplings of eq.~(\ref{cpoddfr}) can be rewritten in the form
%%%%%%%%%%%%%%%%%
\begin{equation}\label{limit}
\lambda_{A\tilde{t}_1\tilde{t}_2} \simeq  -\frac{g_W m_t}{2 M_W} 
\mu  ,  \qquad\qquad
\lambda_{A\tilde{b}_1\tilde{b}_2} \simeq  -\frac{g_W m_b}{2 M_W} 
A_b \tan\beta .
\label{largetanb}
\end{equation}
%%%%%%%%%%%%%%%%%%
That is,
the coupling which is associated with the sbottom pair is 
proportional to $\tan\beta$, so that, eventually, the total
$\tilde{b}_1\tilde{b}_2^*A$ 
cross section will grow with $\tan^2\beta$
while the coupling related to 
the stop pair takes on constant values. In the 
latter, the enhancement of the $\tilde{t}_1\tilde{t}_2^*A$
cross section with $\tan\beta$ is
rather a phase space effect since, as $\tan\beta$ increases, the 
 CP-odd Higgs boson mass decreases considerably 
(the squark masses changing much less instead),
as we can see from Tab.~II. Of course the same is valid in
the former case and that is why our Figs. 
\ref{fig:trtlAtanb}--\ref{fig:brblAtanb} follow the pattern
$\sigma (gg\rightarrow
\tilde{b}_1 \tilde{b}_2^* A )
 \gsim \sigma (gg\rightarrow \tilde{t}_1
 \tilde{t}_2^* A )$ at large $\tan\beta$.

%In other terms, both couplings are proportional to $\tan\beta$, modulus a
%prefactor $\mu$ or $A_b$. Understandably then, the rates for processes
%(\ref{proc}) grow fast with increasing values of the VEV of the Higgs
%fields. In particular, in view of eq.~(\ref{limit}), it is easy to follow
%the pattern $\sigma (gg\rightarrow
%\tilde{b}_1 \tilde{b}_2^* A ) \gsim \sigma (gg\rightarrow \tilde{t}_1
% \tilde{t}_2^* A )$ at large $\tan\beta$: 
%simply because $A_b \gsim \mu$ in that region.
 
Despite of the abundance of 
$\tilde{t}_1\tilde{t}_2^* A$ and $\tilde{b}_1\tilde{b}_2^* A$
events at large $\tan\beta$, overwhelming contributions involving 
the light stop ${\tilde t}_1$ and light Higgs scalar $h$, i.e.,
$\tilde{t}_1\tilde{t}_1^* h$ and, particularly,
$\tilde{t}_1\tilde{t}_2^* h$ (see Tab.~I), would however dominate the 
squark-squark-Higgs production phenomenology. Therefore, it might
seem at first glance that reactions (\ref{proc}) cannot possibly
be disentangled, further considering that at large $\tan\beta$ the
dominant decay modes of both $h$ and $A$ scalars are into $b\bar b$
pairs \cite{ioejames}. This need not to be true though. In fact, 
the reader should recall two important aspects.
Firstly, the lighter scalar quark mass, 
$m_{\tilde{t}_1}$, will most likely be known well before
a statistically significative sample of events of the form (\ref{allproc})
can be collected. It follows that its knowledge can be exploited to remove
$gg \ar {\tilde{q}}_{\chi} {\tilde{q}}^{'*}_{\chi'} \Phi$
candidates with exactly two 
light stop squarks, thus also unwanted
$\tilde{t}_1\tilde{t}_1^* h$ final states.
Secondly, in the $b\bar b$ channel, one should expect experimental
 mass resolutions
to be smaller than the typical mass differences $m_A-m_h \gsim 20$ GeV seen 
for $\tan\beta\lsim35$ (see Tab.~II) \cite{progress}\footnote{The intrinsic 
width $\Gamma_\Phi$
of the $h$ and $A$ Higgs bosons is in that mass range 
about 2.5 GeV at the most.}. Needless to say, the light scalar $h$ 
ought to have been discovered (and $m_h$ measured)
by then, for the sake of the all SUSY theory, so that a suitable 
selection of $b\bar b$ pairs far away from the $m_h$ resonance
(or, at worse, an event counting operation in the case of overlapping 
$h$ and $A$ mass peaks, at extremely large $\tan\beta$) 
would aid to reduce $\tilde{t}_1\tilde{t}_2^* h$ events also.

As for the low $\tan\beta$ region, as
intimated a few paragraphs above, we can appreciate in 
Fig.~\ref{fig:trtlAtanb}
the beneficial effect of a large and negative value of $A_0$, 
in terms of  the $\sign(\mu)$ dependence of the $A$ production rates.
For example, for $\tan\beta\sim 2$, the cross sections for $\mu>0$ and
$\mu<0$ take very different values, by an order of magnitude.
In fact, one has that, for $A_0\lsim -300$ GeV,  
%%%%%%%%%%%%%%%%
\begin{equation}
\sigma (gg\rightarrow \tilde{t}_1  \tilde{t}_2^* A ) \sim 10^{-3} 
~{\mathrm{pb}}\;\;\;
{\rm for }\;\;\; \mu > 0, 
\qquad
\sigma (gg\rightarrow \tilde{t}_1  \tilde{t}_2^* A ) \sim 10^{-4} 
~{\mathrm{pb}} \;\;\;
{\rm for }\;\;\; \mu < 0\;.
\end{equation}
%%%%%%%%%%%%%%%%%%
Therefore, in this scenario one might aim to constrain the actual value of
sign($\mu$)  from the
 $\tilde{t}_1  \tilde{t}_2^* A$ final states alone, 
given such a large difference.
Unfortunately, the total number of events (after having considered
the decay rates, the finite efficiency and resolution of experimental 
analyses, etc.) is again not so large, so that one would presumably be better
off by relying on reaction $ gg\rightarrow \tilde{t}_1  \tilde{t}_2^* h$.
In this respect though, one thing is worth spotting, i.e., the much larger
value of $m_A$ as compared to $m_h$ if $\tan\beta$ is small, see
 Tab.~II. A consequence of this is that the decay
patterns of the two Higgs bosons are very different. Whereas the
light one would only decay into $b\bar b$ pairs, the pseudoscalar one
would mainly yield $t\bar t$ pairs \cite{ioejames}.
Given the huge QCD noise of the LHC, the latter might in the end become 
a competitive approach, especially if a clean electron/muon tag can be 
achieved in the (anti)top decays. 

But, let us now turn our attention to the other strong dependence of
the production rates of $gg\ar \tilde{t}_1 \tilde{t}_2^* A$ and 
$gg\ar \tilde{b}_1 \tilde{b}_2^* A$:  the one on 
the common trilinear coupling $A_0$. This is in fact
the most noticeable feature of both Figs.
\ref{fig:trtlAtanb}--\ref{fig:brblAtanb}: that the sensitivity
to $A_0$ of the production cross sections
provides the unique
possibility of constraining, possibly the sign, and hopefully the
magnitude, of this fundamental M-SUGRA parameter. 
Indeed, we have come to believe 
that this is the main novelty that should be attributed to the
phenomenological potential of the processes we are studying, as one 
might quite rightly expect that the determination
of  $\tan\beta$ will come first from studies in the pure
Higgs sector (i.e., via SM-like Higgs production and decay mechanisms), 
especially considering the theoretical upper limit on $m_h$. Should this be 
the case, far from overshadowing the usefulness of reactions (\ref{proc}),
the knowledge of $\tan\beta$ would further help to constrain $A_0$.
Let us see how.

For a start, to observe by the thousand events involving 
$A$-production with pairs of stop quarks would
induce the following reasoning:
%%%%%%%%%%%%%%%%%
\begin{eqnarray}
\sigma (gg\rightarrow \tilde{t}_1  \tilde{t}_2^* A ) \gsim  10^{-2} \; 
{\mathrm{pb}}
\Rightarrow \left \{ \begin{array}{c}  -500 < A_0 < 0 \;\;
{\rm for} \;\; 22 \lsim \tan\beta \lsim 40 \;\;{\rm and}
\;\; \mu=\pm.  \\[3mm] \;\;\,  0 \le A_0 < 500 \; \; 
{\rm for } \;\; 40 \lsim \tan\beta \lsim 42 \;\; {\rm and}\;\; \mu=\pm.
 \end{array} \right. \;
\end{eqnarray}
%%%%%%%%%%%%%%%%%
That is to say, unless $40\lsim \tan\beta\lsim 42$ (a very small
corner of the M-SUGRA
parameter space), to observe such a rate of $\tilde{t}_1  \tilde{t}_2^* A$
events would mean that $A_0$ is necessarily negative (whichever the sign
of $\mu$).

Incidentally, we would like the reader to spot in 
Fig.~\ref{fig:trtlAtanb} that are the lines corresponding to $A_0=300$ GeV
(denoted by the arrow)  those stretching to the far right of the plot, thus
inverting the trend of decreasing rates with growing $A_0$.
In other terms, such curves  represent a true
lower limit on the value of this cross section
(practically for all $\tan\beta$ values), so that the latter is
bound to be in the range
%%%%%%%%%%%%%%%%%
\begin{eqnarray}
10^{-4}~{\mathrm{pb}}\lsim \sigma (gg\rightarrow \tilde{t}_1  \tilde{t}_2^* A) 
\lsim 10^{-1} \; {\mathrm{pb}} \;,
\end{eqnarray}
%%%%%%%%%%%%%%%
values well within the reach of the LHC luminosity !

Similarly, one can proceed to analyse 
$gg\rightarrow \tilde{b}_1  \tilde{b}_2^* A$ from Fig.~\ref{fig:brblAtanb}.
Schematically,
%%%%%%%%%%%%%%%%%
\begin{eqnarray}
\sigma (gg\rightarrow \tilde{b}_1  \tilde{b}_2^* A ) \gsim  10^{-2} \; 
{\mathrm{pb}}
\Rightarrow \left \{ \begin{array}{c}  -500 < A_0 < 0 \;\;
{\rm for} \;\; 28 \lsim \tan\beta \lsim 40 \;\;{\rm and}
\;\; \mu=\pm.  \\[3mm] \;\;\,  0 \le A_0 < 500 \; \; 
{\rm for } \;\; 40 \lsim \tan\beta \lsim 42 \;\; {\rm and}\;\; \mu=\pm.
 \end{array} \right. \;
\end{eqnarray}
%%%%%%%%%%%%%%%%%
Once again, to observe  $\tilde{b}_1  \tilde{b}_2^* A$ signals at such
a rate would
force $A_0$ to be negative over most of the M-SUGRA parameter space.

As for peculiar trends in Fig.~\ref{fig:brblAtanb}, two behaviours
worth commenting on are the following. Firstly, that the production
rates decrease with diminishing $\tan\beta$ much more than they do
in case of stop production, particularly if $\tan\beta \lsim 30$.
Secondly, that the cross sections exactly vanish in the case 
$A_0=500$ GeV, when $\tan\beta=24(27)$ if $\mu>0(\mu<0)$, 
as induced by the $\propto(\mu-A_b \tan\beta)$
 dependence of the production vertex, when $|\mu|\ll | A_b \tan\beta|$
and $A_b$ changes its sign.

Another aspect made clear by both these two figures is
that current experimental bounds tend to exclude 
only extreme parameter regions, i.e.,
where $A_0$ is strongly negative and/or where
$\tan\beta$ is extremely high. On the one hand, LEP2 
has almost exhausted its SUSY discovery potential, as most
of the data have already been collected and/or analysed, whereas
at Tevatron, the present $m_{\tilde{t}_1}\OOrd120$ GeV limit
on the lighter stop mass is unlikely to be increased by the 
new runs to the typical values of Tab.~II.
On the other hand, the bulk of the $(A_0,\tan\beta,\sign(\mu))$ parameter
space investigated here, where processes (\ref{proc})  could well
be detected and studied at the LHC,
appears in Figs.~\ref{fig:trtlAtanb}--\ref{fig:brblAtanb}
far beyond the reach of the present colliders.
Therefore, in the very short term, one should not expect that new experimental
limits can modify drastically the look of our plots. In particular, notice 
that the presence of ${\tilde t}_2$
and ${\tilde b}_2$ squarks in the final state of processes (\ref{proc})
implies that the corresponding production rates at Tevatron 
are negligible, even for optimistic luminosities,
because of the enormous phase space suppression 
(see Tab.~II)\footnote{The gluon luminosity is much poorer too, as compared
to the CERN hadron collider.}.
Therefore, we believe that, when the LHC will start running, 
most of the M-SUGRA parameter space discussed here will still be unexplored.

Bringing together the various results obtained so far on $A_0$, $\tan\beta$
and $\sign(\mu)$, we attempt to summarise our findings in 
Tab.~III.
There, we list the restrictions that can in principle
be deduced on the above three
parameters by studying the two processes (\ref{proc}), assuming
that none of these quantities is known beforehand. Indeed, an enormous 
area of the M-SUGRA space can be put under scrutiny,
particularly involving $A_0$ and $\tan\beta$. 
The prospect of the latter quantity being already known 
by the time $gg\ar \tilde{t}_1  \tilde{t}_2^* A$ and
$gg\ar \tilde{b}_1  \tilde{b}_2^* A$ studies begin
would be even more  exciting. In such a case, a vertical line
could be drawn in Figs.~\ref{fig:trtlAtanb}--\ref{fig:brblAtanb}, so that
an accurate measurement of the production cross sections of processes
(\ref{proc}) would precisely pin-point the actual value of $A_0$.

Before closing, we study the dependence of pseudoscalar Higgs
boson production in association with stop and sbottom squarks on
the last two M-SUGRA independent parameters: $M_0$ and $M_{1/2}$.
The main effect of changing the latter is
onto the masses of the final state scalars, through the phase space volume
as well as via propagator effects in the scattering amplitudes.
In other terms, to increase one or the other depletes the
typical cross sections of (\ref{allproc}), simply because
the values of all $m_{\tilde{q}_\chi}$ and $m_\Phi$ get larger. 
Tab.~IV samples such a trend on four among
the dominant production channels, including our two  reactions (\ref{proc}).
As an example, notice that, for  $M_0=300$ GeV and $M_{1/2}=250$ GeV, 
all the squark masses are of the order $\gsim  460$ GeV, whereas for the
heavy Higgs bosons one has that typical values are $\gsim 290$ GeV.
Not surprisingly then, among the processes in Tab.~IV, for such
high $M_0$ and $M_{1/2}$ values, the only ones to survive are those involving
both the lightest squark (i.e., $\tilde{t}_1$) 
and the $h$ scalar 
(for which one necessarily has that $m_h\lsim130$ GeV)
\cite{progress,abdel}. In comparison, 
processes (\ref{proc}) are generally suppressed,
as one heavy mass $m_{{\tilde q}_2}$ is
always present in the final states and since $m_A\gsim m_h$. Therefore,
this last exercise shows that only light $M_0$ and $M_{1/2}$ masses
(say, below 200 and 150 GeV, respectively) 
would possibly allow for pseudoscalar production to be detectable at the LHC.

%%%%%%%%%%%%%%%%%%%%%%
\begin{center}
%%%%%%%%%%%%%%%%%%
\begin{eqnarray}
\begin{array}{|c|c|c|c|c|}\hline
\sigma (gg\rightarrow \tilde{t}_1  \tilde{t}_2^* A ) \; {\mathrm{(pb)}}&
\sigma (gg\rightarrow \tilde{b}_1  \tilde{b}_2^* A ) \; {\mathrm{(pb)}}&
A_0 \; {\mathrm{(GeV)}} & \tan\beta & \sign(\mu) \\ \hline
\gsim 10^{-2} & \gsim 10^{-2} & -500\div 0 & 28\div 40 & \pm \\ \hline
\gsim 10^{-2} & \gsim 10^{-2} & \gsim 0 & 40 \div 42 & \pm \\ \hline
\gsim 10^{-2} & \lsim 10^{-2} & -500 \div -300 & 22 \div 28 & \pm \\ \hline
\lsim 10^{-2} & \gsim 10^{-2} & -300\div 0 & 28 \div 40 & \pm \\ \hline
\lsim 10^{-2} & \gsim 10^{-2} &  \gsim 0 & 40 \div 42 & \pm \\ \hline
\lsim 10^{-2} & \lsim 10^{-2} &  \gsim -500 & \lsim 40 & \pm \\ \hline
\sim 10^{-3} & \lsim 10^{-6} &  \gsim -500 & 2\div 3 & + \\ \hline
\sim 10^{-4} & \lsim 10^{-6} &  \gsim -500 & 2\div 3 & - \\ \hline
\end{array}
\nonumber
\end{eqnarray}
%%%%%%%%%%%%%%%%%
{Table III: Possible restrictions on three 
M-SUGRA parameters derivable from studies of
CP-odd Higgs boson production in association with stop and sbottom squarks.}
\end{center}

%%%%%%%%%%%%%%%%%%%%%%%%%%%%%
\begin{center}
%%%%%%%%%%%%%%%%%%%%%%%
\begin{eqnarray}
\begin{array}{|c|c|c|c|c|c|}\hline
M_0 ({\mathrm{GeV}}) & M_{1/2} ({\mathrm{GeV}}) & 
  \sigma (gg\rightarrow \tilde{t}_1  \tilde{t}_1^* h )
& \sigma (gg\rightarrow \tilde{t}_1  \tilde{t}_2^* h ) &
  \sigma (gg\rightarrow \tilde{t}_1  \tilde{t}_2^* A ) &
  \sigma (gg\rightarrow \tilde{b}_1  \tilde{b}_2^* A ) \\ \hline
200 & 125 & 6.8 \times 10^{-3} & 6.9 \times 10^{-2} & 4.0 \times 10^{-3} &
5.2 \times 10^{-3} \\
200 & 150 & 3.4 \times 10^{-3} & 4.7 \times 10^{-2} & 1.4 \times 10^{-3} &
2.4 \times 10^{-3} \\
200 & 200 & 9.7 \times 10^{-4} & 1.8 \times 10^{-2} & 3.1 \times 10^{-4} &
6.0 \times 10^{-4} \\
300 & 250 & 2.6 \times 10^{-4} & 5.9 \times 10^{-3} & 4.8\times 10^{-5} &
9.8 \times 10^{-5} \\ \hline 
\end{array}
\nonumber
\end{eqnarray}
%%%%%%%%%%%%%%%%%%%%%%%%
{Table IV: 
The variation of the  most significant cross sections (in picobarns)
of processes (\ref{allproc}) with $M_0$ and
 $M_{1/2}$. For reference, the other three M-SUGRA parameters are fixed 
as follows: $A_0=0$, $\tan\beta=40$ and $\sign(\mu)=-$.}
\end{center}

\section{Summary and conclusions}
\label{sec:conclusions}

In summary, we have studied pseudoscalar Higgs boson
production in association with  stop and sbottom squarks at the LHC, in the
context  of the SUGRA inspired MSSM. Our interest in such reactions
was driven by the fact that the squark-squark-Higgs vertices involved,
other than carrying a strong dependence on three free inputs of such a model,
i.e., $A_0$, $\tan\beta$ and $\sign(\mu)$, are not affected by the
presence of additional unconstrained parameters describing
the mixing between physical and chiral squark eigenstates.

We have found that the cross sections of such processes might be  
detectable both at low and high
collider luminosity for not too small values of $\tan\beta$. Indeed,
their production rates are strongly sensitive to the ratio 
the VEVs of the Higgs fields, thus possibly
allowing one to put potent constraints
on such a crucial parameter of the MSSM Higgs sector. 
Furthermore, also the trilinear
coupling $A_0$ intervenes in these events, in such a 
way that visible rates could mainly be possible if this other fundamental
M-SUGRA input is negative. (Indeed, to know the actual value
of $\tan\beta$ from other sources would further help to assess the
magnitude of $A_0$.) As for the sign of the Higgsino mass term,
$\sign(\mu)$, it only marginally affects the phenomenology of such events.
Finally,
concerning the remaining two parameters (apart from mixing effects) of
the M-SUGRA scenario, i.e., $M_0$ and $M_{1/2}$,
it must be said that their values should be such that 
they guarantee a rather light squark and Higgs mass spectrum, in order
the latter to be within the reach of the LHC.

In conclusion, we believe these processes to be potentially very  helpful
in putting stringent limits on several M-SUGRA parameters and 
we thus recommend that their subsequent decay and hadronic dynamics
is further investigated in the context of dedicated experimental simulations, 
which were clearly  beyond the scope of this short letter. As a matter
of fact, of all possible (eighteen in total) squark-squark-Higgs
production modes, involving sbottoms, stops and all Higgs 
mass eigenstates, we
have verified that those including the pseudoscalar
particle are always among
the dominant ones, so that one should not expect the presence of the
others to dash away the hope of detecting and investigating the former.
In this respect, the most competing ones are those involving 
the lightest of the Higgs scalars. 
\begin{center}
\begin{eqnarray}
\begin{array}{cc}   

%\begin{eqnarray}
\begin{array}{|ccc|}\hline
{\rm Particle} & {\mathrm{BR}} 
& {\rm Decay} \\ \hline   
\tilde{t}_1 & \stackrel{96\% 
}{\rightarrow}& \chi_1^+ b \\ \hline
\tilde{t}_2 & \stackrel{44\%
}{\rightarrow}& \chi_2^+ b \\
 & \stackrel{30\%
}{\rightarrow}  & \chi_1^+ b \\
 & \stackrel{16\%
}{\rightarrow}  & {\tilde{b}}_1 W^+ \\ 
 & \stackrel{5\%
}{\rightarrow}  & {\tilde{t}}_1 Z \\ \hline

\tilde{b}_1 & \stackrel{51\%
}{\rightarrow}& \chi_2^0 b \\
 & \stackrel{41\%
}{\rightarrow}  & \chi_1^0 b \\ \hline
\tilde{b}_2 & \stackrel{41\%
}{\rightarrow}& \chi_3^0 b \\
 & \stackrel{32\%
}{\rightarrow}  & \chi_4^0 b \\ 
 & \stackrel{17\%
}{\rightarrow}  & \chi_2^0 b \\ \hline

h & \stackrel{79\%
}{\rightarrow}& b \bar{b} \\
 & \stackrel{12\%
}{\rightarrow}  & \chi_1^0 \chi_1^0 \\
% & \stackrel{5\%
%}{\rightarrow}  & \tau^+ \tau^- \\ 
\hline

H & \stackrel{93\%
}{\rightarrow}& b \bar{b} \\
 & \stackrel{6\%
}{\rightarrow}  & \tau^+ \tau^-  \\ \hline

A & \stackrel{92\%
}{\rightarrow}& b \bar{b} \\
 & \stackrel{6\%
}{\rightarrow}  & \tau^+ \tau^- \\ \hline

H^\pm & \stackrel{75\%
}{\rightarrow} & \tau^\pm \nu \\
 & \stackrel{20\%
}{\rightarrow}  & \chi_1^0 \chi_1^\pm \\ \hline

\end{array}
%\nonumber
%\end{eqnarray}
%%%%%%%%%%%%%%%%%%%%%%%%
&
%%%%%%%%%%%%%%%%%%%%%%%%
%\begin{eqnarray}
\begin{array}{|ccc|}\hline
{\rm Particle} & {\mathrm{BR}} 
& {\rm Decay} \\ \hline   
\chi_1^+ & \stackrel{32\%
}{\rightarrow} & \chi_1^0 u \bar{d}^{\dagger} \\
& \stackrel{32\%
}{\rightarrow} & \chi_1^0 c \bar{s}^{\dagger} \\
& \stackrel{14\%
}{\rightarrow} & \chi_1^0  \tau^+ \nu^{\dagger} \\
& \stackrel{10\%
}{\rightarrow} & \chi_1^0 e^+ \nu^{\dagger} \\
& \stackrel{10\%
}{\rightarrow} & \chi_1^0 \mu^+ \nu^{\dagger} \\ \hline 

\chi_2^+ & \stackrel{47\%
}{\rightarrow} & \chi_2^0 W^+ \\
& \stackrel{24\%
}{\rightarrow} & \chi_1^+ Z \\
& \stackrel{10\%
}{\rightarrow} & \chi_1^+ h \\ 
& \stackrel{9\%
}{\rightarrow} & \chi_1^+ A \\ \hline

\chi_2^0 & \stackrel{69\%
}{\rightarrow} & \chi_1^0 b \bar{b}^{\ddagger} \\ 
& \stackrel{10\%
}{\rightarrow} & \chi_1^0 \tau^+ \tau^- \\ \hline 

\chi_3^0 & \stackrel{35\%
}{\rightarrow} & \chi_1^+  W^- \\ 
& \stackrel{35\%
}{\rightarrow} & \chi_1^-  W^+ \\ 
& \stackrel{15\%
}{\rightarrow} & \chi_1^0  Z \\ \hline

\chi_4^0 & \stackrel{39\%
}{\rightarrow} & \chi_1^+  W^- \\ 
& \stackrel{39\%
}{\rightarrow} & \chi_1^-  W^+ \\ 
& \stackrel{6\%
}{\rightarrow} & \chi_1^0  h \\ 
& \stackrel{6\%
}{\rightarrow} & \chi_1^0  Z \\ \hline

\end{array}

\end{array}
\nonumber
\end{eqnarray}
{Table V: Dominant decay channels and branching ratios (BRs)
of final state (s)particles in (\ref{allproc}),
for $M_0=200$ GeV, 
$M_{1/2}=100$ GeV, $A_0=0$, $\tan\beta=40$ and 
${\mathrm{sign}}(\mu)>0$ \cite{isajet}.}
%\vskip0.25in\noindent
{~~~~~~~~~~~$^\dagger$ \footnotesize{Via off-shell $W^{+}$.}\qquad\qquad
 $^\ddagger$ \footnotesize{Via off-shell $h$.}}
\end{center}
This particle has however a
rather different
decay phenomenology from that of the CP-odd Higgs in most cases, whereas
whenever this is not true, previous knowledge of (stop and sbottom) squark 
and/or Higgs mass values can be of some help,  so that in the end
it should not be difficult to disentangle the two scalars.

For example, let us consider the signal 
$gg\ar {\tilde{t}}_1{\tilde{t}}_2^* A$ and some
possible signatures of it. For the choice of parameters
given in the caption and in the fourth column of Tab.~I, it yields
some 3,000 events per year at the LHC. From Tab.~V, one deduces
that a possible decay chain could be the following:
$$
\arraycolsep=0pt 
\begin{array}{lllllll}\label{chain}
{\tilde{t}_1} && &~~~~{\tilde{t}}_2^* &&A~~~~~ & \\
\downarrow &&& ~~~\downarrow &&\downarrow~~~~~ & \\
\chi_1^+ \,+\,b\; & &&~~~~\chi_1^- \,+\,\bar b &~&b \,+\,\bar b &  \\
\downarrow &&& ~~~\downarrow && \; & \\
q \,+\,\bar q'\,+\,\chi_1^0& && ~~~~\ell^-\,+\,\nu\,+\,\chi^0_1 &&&
\end{array}
$$
in which 
$q\bar q'=u\bar d,c\bar s$ and $\ell=e,\mu,\tau$. Considering also
the charge conjugated $\chi_1^+\chi_1^-$ decays, the final signature
would then be 
`$2~{\mathrm{jets}}~+~4b~+\ell^\pm+{E}_{\mathrm{miss}}$', further
recalling
that the two $\chi^0_1$'s and the neutrino $\nu$ produce
missing energy, ${E}_{\mathrm{miss}}$.
The total BR of such decay sequence  is 0.12 only, so that about 360
squark-squark-Higgs events would survive. One may further assume 
a reduction factor of about 0.25 because of the overall 
efficiency $\varepsilon_b^4$
to tag four displaced vertices (assuming $\varepsilon_b\approx0.7$). This
ultimately yields something less than 100 events per year, a respectable
number indeed. In addition, one should expect most of the signal events
to lie in the detector acceptance region, since leptons and jets
originate from decays of heavy objects. 

Such a signature has peculiar features that should help
in its selection: a not too large hadronic multiplicity, six jets in total,
each rather energetic (in fact, note that 
$m_{\chi^\pm_1}-m_{\chi^0_1}\approx30$ GeV and 
$m_{{\tilde{t}_1}}, m_{{\tilde{t}_2}}\gg m_{\chi^\pm_1}\approx63$ GeV),  
so that their reconstruction from the detected 
tracks should be reasonably accurate;
a high transverse momentum and isolated lepton to be used as
trigger; large $E_{\mathrm{miss}}$ to reduce non SUSY processes;
four tagged $b$-jets that
can be exploited to suppress the `$W^\pm$ + light jet' background from QCD 
with one $b\bar b$ pair resonating at the $A$ mass 
(which is well above the $h$ one: see Tab.~II and recall the discussion in 
Sect.~\ref{sec:results} about the interplay between 
$gg\ar {\tilde{t}_1}{\tilde{t}}_2^*A$ and 
$gg\ar {\tilde{t}_1}{\tilde{t}}_2^*h$ events). Even the 
background from $gg\ar {\tilde{t}_1}{\tilde{t}_2^*}Z$ events, 
with $Z\ar b\bar b$, 
potentially very dangerous because
irreducible and since $m_{{\tilde{t}_2}}-m_{{\tilde{t}_1}}\gg M_Z$
 (Tab.~II),
should easily be dealt with. In fact,
notice that $m_A\gg M_Z$, so that to select only events
for which  
$M_{bb}\not\approx M_Z$ would presumably allow one to reduce also such a
noise to manageable levels. 

Notice that the M-SUGRA point just discussed corresponds to a rather
low lightest chargino mass, though still roughly consistent with the latest
bounds drawn by the Particle Data Group (PDG) \cite{PDG}. However,
preliminary results from LEP and Tevatron have meanwhile increased the limit on
$m_{\chi_1^\pm}$, up to 80--90 GeV or so. Thus, we also have considered a
second parameter combination yielding sizable production rates, but now
satisfying the latter constraint: e.g.,
that in the first line of Tab.~IV (see the caption for the parameter
setup), for which $m_{\chi_1^\pm}=89$ GeV, right at the edge of
the exclusion band. Considering again 
`$2~{\mathrm{jets}}~+~4b~+\ell^\pm+{E}_{\mathrm{miss}}$' decays, starting
from some 400 signal events every 100 inverse femtobarns produced in 
the $2\ar3$ scattering (fifth column in Tab.~IV), one ends up with 12 
events, as the total decay BR is basically the same as before and including
again the factor $\varepsilon_b^4$. The number is reduced by an order 
of magnitude, but still sizable.

As a matter of fact, other signatures can possibly  be even more accessible.
Let us now take, e.g., $M_0=M_{1/2}=125(130)[140]$ GeV 
(with again $\tan\beta=40$, $A_0=0$ and sign$(\mu)=-$).
For such settings, the lightest chargino mass is $m_{\chi_1^\pm}=90(93)[101]$
GeV. In correspondence, one gets $\sigma(gg\ar
{\tilde{t}_1}{\tilde{t}}_2^*A)=96(79)[39]$ fb, i.e., some
9,600(7,900)[3,900] events per luminosity year. For these last three 
combinations of M-SUGRA parameters, it turns out that an 
interesting decay sequence could be the following: 
$$
\arraycolsep=0pt 
\begin{array}{lllllll}\label{newchain}
{\tilde{t}_1} && &~~~~{\tilde{t}}_2^* &&A~~~~~ & \\
\downarrow &&& ~~~\downarrow &&\downarrow~~~~~ & \\
\chi_1^+ \,+\,b\; & &&~~~~\chi_1^- \,+\,\bar b &~&b \,+\,\bar b &  \\
\downarrow &&& ~~~\downarrow && \; & \\
{\tilde{\tau}}_1^+ \,+\,\nu\,+\,b& && ~~~~ 
{\tilde{\tau}}_1^- \,+\,\bar\nu\,+\,\bar b &&& \\
\downarrow &&& ~~~\downarrow && \; & \\
\tau^+~+~\nu\,+\,b\,+\,\chi_1^0 & &&
 ~~~~\tau^-\,+\,\bar\nu~+~\bar b\,+\,\chi^0_1 &&&
\end{array}
$$
Apart from the BR of the channel ${\tilde{t}}_2^*\ar\chi_1^-\bar b$,
which ranges at 25\%, the others are
all dominant and close to unity. Here,
the final signature is
`$4b~+~\tau^+\tau^-~+~E_{\mathrm{miss}}$' with a total BR of about
23\% in all cases. Thus, after multiplying by $\varepsilon_b^4$,
 one finally gets 528(454)[218] detectable events every 100 fb$^{-1}$. 

This additional channel appears particularly neat thanks to an even
smaller jet multiplicity. In addition, all such jets should 
be  rather energetic,
as $m_{\chi_1^\pm}-m_{\tilde{\tau}}\approx19(17)[15]$ GeV and
   $m_{\tilde{\tau}}-m_{\chi_1^0}  \approx21(25)[31]$ GeV. 
Standard backgrounds from `$Z$~+~jet' production could be strongly
suppressed because  of the absence of light-quark jets and the
presence of four heavy ones. From one pair of these, one could 
further attempt
to reconstruct the $A$ mass, at around 114(120)[135] GeV. 
Finally, the large amount of $E_{\mathrm{miss}}$ building up 
because of the four neutrinos and two neutralinos could prove to be
a further good handle against non-SUSY processes. As for irreducible
SUSY backgrounds, notice the poor decay rate 
BR$({\tilde{t}}_2^*\ar{\tilde{t}}_1^* Z)\approx7\%$ ! (Typical 
stop masses are around 380(388)[406] and 240(248)[265] and for
${\tilde{t}}_2$ and  ${\tilde{t}}_1$, respectively.)
 
These are just a few illustrative examples of some possible manifestations
of squark-squark-Higgs events at the LHC. Dedicated 
analyses of all  mechanisms of the form
$gg \ar {\tilde{q}}_{\chi} {\tilde{q}}^{'*}_{\chi'} \Phi$,
for any possible combination
of $q^{(')}=t,b$, $\chi^{(')}=1,2$ and $\Phi=H,h,A,H^\pm$,
of their interplay and a simulation of possible 
backgrounds and detection strategies is now under way \cite{progress}.
\vskip0.25cm\noindent{\bf Acknowledgements}~ 
%\section*{Acknowledgements}
%
We thank H. Dreiner for useful discussions. 
S.M. acknowledges the financial support from the UK PPARC,
A.D. that from the Marie Curie Research Training Grant
ERB-FMBI-CT98-3438.

\vfill\clearpage

%

%%%%%%%%%%%%%%%%%%%%%
\begin{figure}
\centerline{\epsfig
{figure=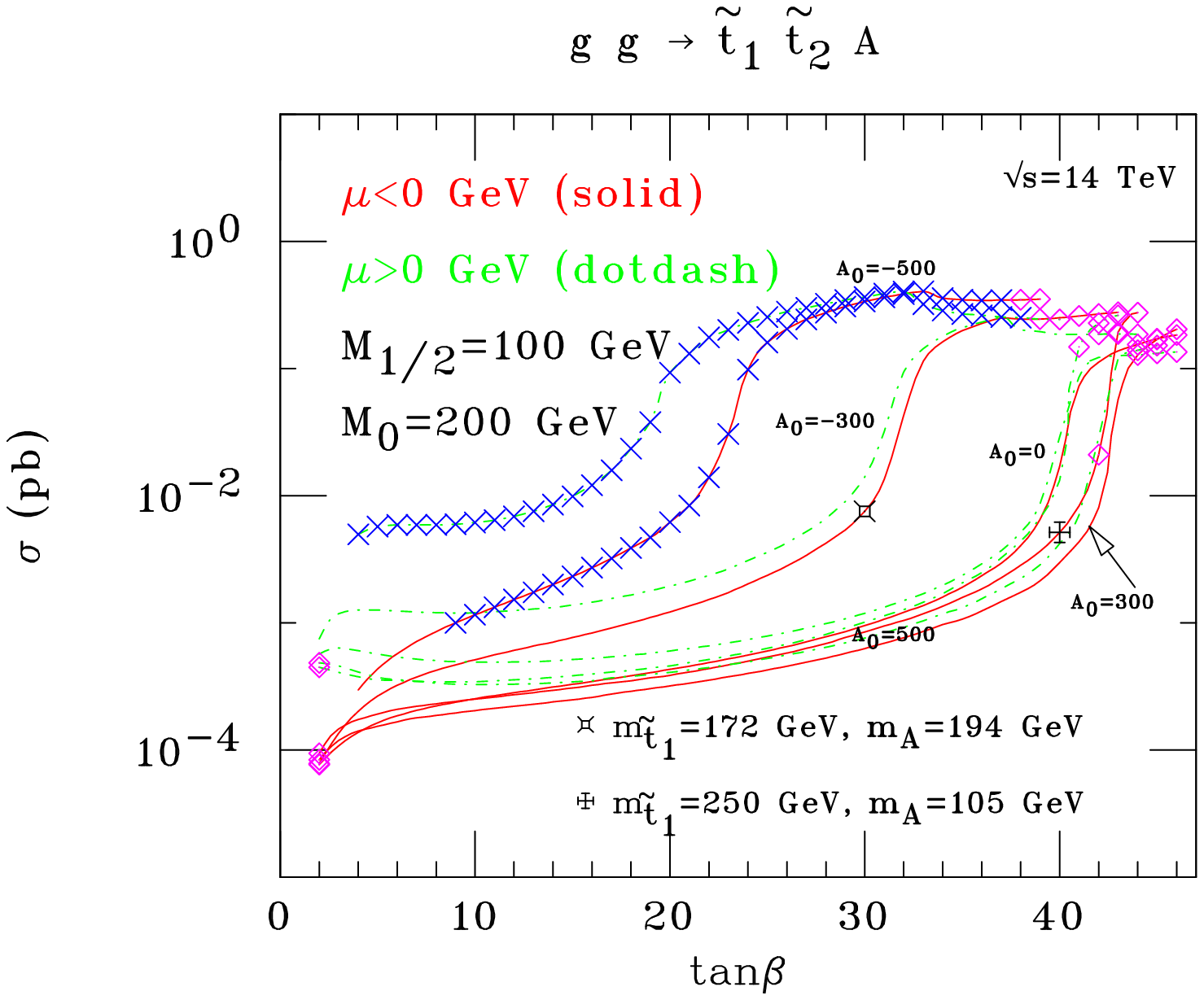,angle=90}}
\caption{Total cross sections for process $gg\rightarrow \tilde{t}_1
\tilde{t}_2^* A$ as a function of $\tan\beta$. These are plotted for
both positive (dotdash) and negative (solid) values of $\mu$ and for
a selection of  $A_0$ values.
(The other M-SUGRA
parameters have been chosen as $M_0=200 $ GeV and $M_{1/2}=100$ GeV.)
The symbol `` $\diamond$ ($\times$) ''
 is used to indicate data points forbidden 
by experimental bounds from 
direct searches of the lightest Higgs(squark) scalar,
yielding $2\lsim \tan\beta \lsim 40$($m_{{\tilde t}_1}>120$ GeV).
Two points with the masses of the
lightest stop and the CP-odd Higgs are also shown.  }
\label{fig:trtlAtanb}
\end{figure}
%%%%%%%%%%%%%%%%%%%%%

%%%%%%%%%%%%%%%%%%%%%
\begin{figure}
\centerline{\epsfig{figure=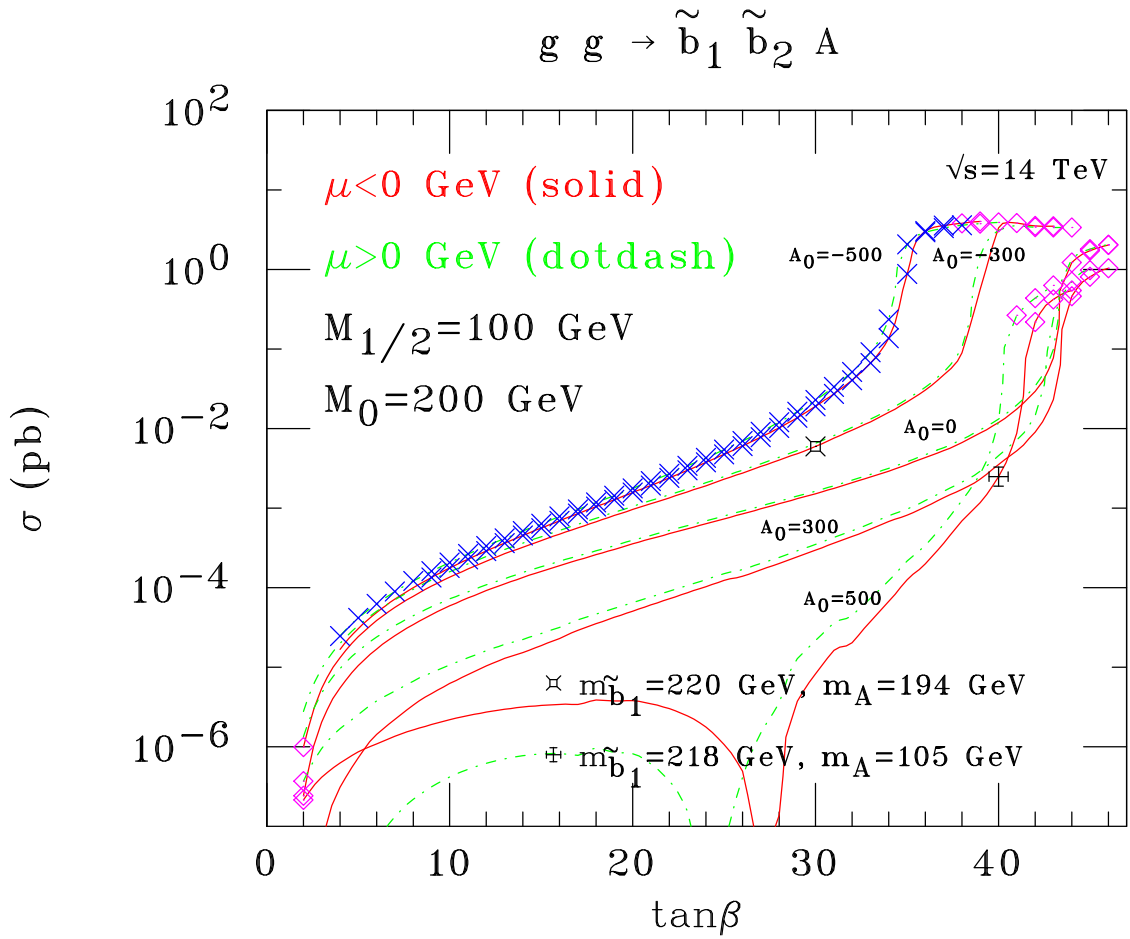,angle=90}}
\caption{Same as in Fig.~\ref{fig:trtlAtanb} for process $gg\rightarrow
 \tilde{b}_1\tilde{b}_2^* A$.}
\label{fig:brblAtanb}
\end{figure}
%%%%%%%%%%%%%%%%%%%%%%

\end{document}